\documentclass[english,aps,preprint]{revtex4}
\begin{document}

\title{Generalized Wick's theorem at finite temperature for a quadratic Hamiltonian}

\author{M O C Pires \\ Centro de Ci\^encias Naturais e Humanas, \\ Universidade Federal
do ABC,\\ Rua Santa Ad\'elia, 166. Santo Andr\'e - SP - Brazil  CEP 09.210-170}

\begin{abstract}
In reference \cite{gau60}, Michel Gaudin showed the Wick's theorem
at finite temperature using a diagonal Hamiltonian. We extend the Gaudin's prove for a statistical density operator which depend
on a quadratic Hamiltonian. To illustrate the utility of the theorem, we evaluate the ratio $\sqrt{[\langle\hat{N}^2\rangle-\langle\hat{N}\rangle^2]/\langle \hat{N}\rangle}$ of a homogeneous
weakly interacting Bose gas at temperature below the Bose-Einstein condensation temperature. At this condition,  the quadratic Hamiltonian approximation is valued and, in this evaluation, 
we show the sub-Poissonian behaviour of the fundamental state distribution
at zero temperature.  
\end{abstract}

\maketitle

\section{Introduction}
\label{session1}

We are interested in describing a quantum system enclosed in a volume $V$ liable to the energy and particle reservoir. Such system branches of science as quantum gases and liquid, nuclear models, quantum information and cosmological dark matter. Attention to these systems is primarily due to the possibility of a deeper understanding of quantum thermodynamics in open systems.
 
At zero temperature,  the system can be represented  by a symmetric (regarding to boson particle) or antissymmetric (regarding to fermion particles) state vector, $|\Psi(t)\rangle$,  in the Fock space. The dynamics of the system is achieved by the solution of the Schrodinger equation, $ |\Psi(t)\rangle=e^{-i\frac{\hat{H}t}{\hbar}}|\Psi\rangle$, where $\hat{H}$ is the Hamiltonian operator and $|\Psi\rangle$ is the initial state vector. 

In contact to the energy and particle reservoir, the system has a finite temperature and can be described by  the statistical density operator in the Fock space, $\hat{\cal D}$, whose properties are hermitian, positive defined and the trace equal to unity. Hence, the thermodynamics variables are determined by the ensemble average $\langle \hat{\cal O}\rangle$ where $\hat{\cal O}$ is the quantum observable related to the thermodynamics variable and is obtained by the prescription,
\begin{equation}
\langle \hat{\cal O}\rangle=\mbox{Tr}\hat{\cal D}\hat{\cal O}
\label{1}
\end{equation}
where the trace is over the Fock space.  

In statistical mechanics, the statistical density operator is determined by the principle of maximum entropy which is defined in terms of the statistical operator as the ensemble average,
\begin{equation}
S=-k_B\mbox{Tr}\hat{\cal D}\ln\hat{\cal D}
\end{equation}
where the $k_B$ is the Boltzmann constant. Whereas the system is in contact to the energy and the particle reservoir, we have introduced Lagrange multipliers, $T$ corresponding to the temperature and $\mu$ corresponding to the chemical potential, to constraint the ensemble average energy $\bar{E}$ and the particle number average $\bar{N}$. Hence, we are led to minimize the functional,
\begin{equation}
\phi(\hat{\cal D})=\mbox{Tr}\hat{\cal D}(k_BT\ln\hat{\cal D}+\hat{\cal K})
\end{equation} 
where $\phi$ is the functional called the thermodynamic potential and $\hat{\cal K}=\hat{H}-\mu \hat{N}$ is the canonical Hamiltonian with $\hat{N}$ as the particle number operator. The variational equation for $\hat{\cal D}$ yields the statistical operator as,
\begin{equation}
\hat{\cal D}=\frac{e^{-\beta\hat{\cal K}}}{\mbox{Tr}e^{-\beta\hat{\cal K}}}.
\label{gau01}
\end{equation}
where $\beta=1/k_BT$.

Without loss of generality and for the sake of the simplicity to take the trace in the ensemble average, we build  any operators, $\hat{\cal O}$,  as linear combination of the products of the creation and annihilation operators $\alpha_i^\dagger$ and $\alpha_i$ of the state vectors in the Fock space. Thus we can written $\hat{\cal O}$ as,
\begin{equation}
\hat{\cal O}= \sum_{n,m}U_{nm}(\underbrace{\alpha^\dagger_i\alpha^\dagger_j\cdots\alpha^\dagger_k\alpha^\dagger_l}_n \underbrace{\alpha_i\alpha_j\cdots\alpha_k\alpha_l}_m),
\end{equation}
where $U_{nm}$ is a constant. The operators $\alpha_i^\dagger$ and $\alpha_i$ satisfy the commutation rules,     
\begin{equation}
[\alpha_{i},\alpha_{j}^{\dagger}]_\epsilon\equiv\alpha_{i}\alpha_{j}^{\dagger}-\varepsilon\alpha_{j}^{\dagger}\alpha_{i}=\delta_{ij},
\end{equation}
where $\varepsilon$ equal to $+1$ when $\alpha_{i}^{\dagger}$ creates a boson state and $-1$ when $\alpha_{i}$ creates a fermion state. 

In order to proceed the trace in the average prescription,  it is appropriated to approach the canonical Hamiltonian to the quadratic Hamiltonian \cite{bla86},
\begin{equation}
\hat{\cal K}_{eff}=\frac{1}{2}\sum_{ij}[h_{ij}(\alpha_{i}^{\dagger}\alpha_{j}+\alpha_{i}\alpha_{j}^{\dagger})+\Delta_{ij}^{*}\alpha_{i}\alpha_{j}+\Delta_{ij}\alpha_{i}^{\dagger}\alpha_{j}^{\dagger}],
\label{2}
\end{equation}
where  the elements $h_{ij}$ and $\Delta_{ij}$ are dynamics parameters.

In the quadratic form, the Hamiltonian can be diagonalised by a canonical Bogoliubov transformation
in creation and annihilation operators, $\alpha_{i}^{\dagger}$ and
$\alpha_{i}$, 
\begin{equation}
\cases{
\alpha_{i}^{\dagger}=\sum_{j}u_{ij}\eta_{j}^{\dagger}-v_{ij}\eta_{\bar{j}}\cr
\alpha_{i}=\sum_{j}u_{ij}^{*}\eta_{j}-v_{ij}^{*}\eta_{\bar{j}}^{\dagger},},
\label{gau02}
\end{equation} 
where the sign $\bar{j}$ consist of time-reversed states and the transformation parameters, $u_{ij}$ and $v_{ij}$, obey  the normalization conditions,
\begin{eqnarray}
\sum_{ij}(|u_{ij}|^{2}-\epsilon |v_{ij}|^{2})=1,
\end{eqnarray}
 ensuring the transformation be canonical. Such parameters are given to diagonalize the canonical Hamiltonian and be possible to write the quadratic Hamiltonian, (\ref{gau02}), in a diagonal form, 
\begin{eqnarray} 
\hat{\cal K}_{eff}=\sum_{i}E_{i}\eta_{i}^{\dagger}\eta_{i}.
\end{eqnarray}

The generalized Wick's theorem demostrated in  \cite{gau60} asserts the ensemble averages is equal to the sum over all possible contracted terms,
\begin{eqnarray}
\langle {\cal O} \rangle&=& \sum_{n,m}U_{nm}(\langle\alpha^\dagger_i\alpha^\dagger_j\cdots\alpha^\dagger_k\alpha^\dagger_l \alpha_i\alpha_j\cdots\alpha_k\alpha_l\rangle)\nonumber \\
&=&\sum_{n,m}U_{nm}(\widehat{\alpha^\dagger_i\alpha^\dagger_j}\langle\alpha^\dagger_l\alpha^\dagger_m\cdots\alpha_k\alpha_l\rangle+\epsilon
\widehat{\alpha^\dagger_i\alpha^\dagger_l}\langle\alpha^\dagger_j\alpha^\dagger_m \cdots\alpha_k\alpha_l\rangle+\cdots)
\end{eqnarray}
where the contracted terms are given by,
\begin{eqnarray}
\cases{\widehat{\alpha^\dagger_i\alpha^\dagger_j}=0\cr \widehat{\alpha^\dagger_i\alpha_j}=\frac{\delta_{i,j}}{1-\epsilon e^{\beta E_i}}}.
\end{eqnarray}
Then the ensemble averages can be written as linear combination of ensemble averages of the product of one creation and other annihilation operators. 

In the section \ref{session2}, we demonstrate the generalized Wick's theorem at finite temperature using the quadratic Hamiltonian (\ref{2}). Differently from the generalized Wick's theorem demostrated in  \cite{gau60}, the contracted terms depend on the transformation parameters $u_{ij}$ and $v_{ij}$.  In section \ref{session3} we use this theorem to show the sub-Poissonian distribution in the particle number ensemble average to a homogeneous weakly interacting Bose gas. We conclude the subject in section \ref{session4}.

\section{Demonstration of the Wick's theorem}
\label{session2}

Let a ensemble average be $\langle abc\cdots ef\rangle$ where $a$, $b$ and etc. are creation or anihilation operators. The generalized Wick's theorem asserts,
\begin{eqnarray}
\langle abc\cdots ef\rangle=\widehat{ab}\langle bc\cdots ef\rangle+\epsilon\widehat{ac}\langle bd\cdots ef\rangle+\cdots.
\end{eqnarray}

To proceed the demonstration,  we assume the operator $a$ as
$\alpha_{i}^{\dagger}$ and write the trace in terms of transformed
operators, $\eta_{j}^{\dagger}$ and $\eta_{\bar{j}}$,
\begin{equation}
\mbox{Tr}\{\alpha_{i}^{\dagger}bc\cdots
ef\hat{\cal D}\}=\sum_{j}u_{ij}\mbox{Tr}\{\eta_{j}^{\dagger}bc\cdots
ef\hat{\cal D}\}-v_{ij}\mbox{Tr}\{\eta_{\bar{j}}bc\cdots ef\hat{\cal D}\}.
\label{gau05}
\end{equation}
whereas, $\eta_{j}^{\dagger}\hat{\cal D}=e^{\beta E_{j}}\hat{\cal D}\eta_{j}^{\dagger}$ and  $\eta_{j}\hat{\cal D}=e^{-\beta E_{j}}\hat{\cal D}\eta_{j}$, due to the canonical hamiltonian, $\hat{\cal K}_{eff}$, 
being diagonal in this news operators, the trace depend on, 
\begin{eqnarray}
\mbox{Tr}\{ bc\cdots ef\eta_{j}^{\dagger}\hat{\cal D}\}=e^{\beta
  E_{j}}\mbox{Tr}\{\eta_{j}^{\dagger}bcd\cdots ef\hat{\cal D}\},
\label{gau07}
\end{eqnarray}

Then we have 
\begin{eqnarray}
 &  & \mbox{Tr}\{\eta_{j}^{\dagger}bcd\cdots ef\hat{\cal D}\}=\nonumber \\
 & = & \frac{[\eta_{j}^{\dagger},b]_\epsilon}{(1-\epsilon e^{\beta
 E_{j}})}\mbox{Tr}\{ cd\cdots
 ef\hat{\cal D}\}+\epsilon\frac{[\eta_{j}^{\dagger},c]_\epsilon}{(1-\epsilon e^{\beta
 E_{j}})}\mbox{Tr}\{ bd\cdots ef\hat{\cal D}\}+\nonumber \\
 & + & \frac{[\eta_{j}^{\dagger},d]_\epsilon}{(1-\epsilon e^{\beta
 E_{j}})}\mbox{Tr}\{ bc\cdots
 ef\hat{\cal D}\}+\cdots+\frac{[\eta_{j}^{\dagger},f]_\epsilon}{(1-\epsilon e^{\beta
 E_{j}})}\mbox{Tr}\{ bcd\cdots e\hat{\cal D}\}.
\label{gau08}
\end{eqnarray}
and
\begin{eqnarray}
 &  & \mbox{Tr}\{\eta_{j}bcd\cdots ef\hat{\cal D}\}=\nonumber \\
 & = & \frac{[\eta_{j},b]_\epsilon}{(1-\epsilon e^{-\beta E_{j}})}\mbox{Tr}\{
 cd\cdots ef\hat{\cal D}\}+\epsilon\frac{[\eta_{j},c]_\epsilon}{(1-\epsilon e^{-\beta
 E_{j}})}\mbox{Tr}\{ bd\cdots ef\hat{\cal D}\}+\nonumber \\
 & + & \frac{[\eta_{j},d]_\epsilon}{(1-\epsilon e^{-\beta E_{j}})}\mbox{Tr}\{
 bc\cdots ef\hat{\cal D}\}+\cdots+\frac{[\eta_{j},f]_\epsilon}{(1-\epsilon e^{-\beta
 E_{j}})}\mbox{Tr}\{ bcd\cdots e\hat{\cal D}\}.
\label{gau09}
\end{eqnarray}

Using (\ref{gau08}) and (\ref{gau09}) in (\ref{gau05}), we have 
\begin{eqnarray}
 &  & \mbox{Tr}\{\alpha_{i}^{\dagger}bc\cdots ef\hat{\cal D}\}=\nonumber \\
 & = & \sum_{j}\Bigg(\frac{u_{ij}[\eta_{j}^{\dagger},b]_\epsilon}{(1-\epsilon
 e^{\beta E_{j}})}-\frac{v_{ij}[\eta_{\bar{j}},b]_\epsilon}{(1-\epsilon e^{-\beta
 E_{j}})}\Bigg)\mbox{Tr}\{ cd\cdots ef\hat{\cal D}\}+\nonumber \\
 & + &
 \epsilon\sum_{j}\Bigg(\frac{u_{ij}[\eta_{j}^{\dagger},c]_\epsilon}{(1-\epsilon
 e^{\beta E_{j}})}-\frac{v_{ij}[\eta_{\bar{j}},c]_\epsilon}{(1-\epsilon e^{-\beta
 E_{j}})}\Bigg)\mbox{Tr}\{ bd\cdots ef\hat{\cal D}\}+\cdots\nonumber \\
 & + & \sum_{j}\Bigg(\frac{u_{ij}[\eta_{j}^{\dagger},f]_\epsilon}{(1-\epsilon
 e^{\beta E_{j}})}-\frac{v_{ij}[\eta_{\bar{j}},f]_\epsilon}{(1-\epsilon e^{-\beta
 E_{j}})}\Bigg)\mbox{Tr}\{ bcd\cdots e\hat{\cal D}\},
\label{gau10}
\end{eqnarray}
in the same way using $a=\alpha_{i}$, we have 
\begin{eqnarray}
 &  & \mbox{Tr}\{\alpha_{i}bc\cdots ef\hat{\cal D}\}=\nonumber \\
 & = & \sum_{j}\Bigg(\frac{[\eta_{j},b]_\epsilon u_{ij}^{*}}{(1-\epsilon
 e^{-\beta
 E_{j}})}-\frac{[\eta_{\bar{j}}^{\dagger},b]_\epsilon v_{ij}^{*}}{(1-\epsilon e^{\beta
 E_{j}})}\Bigg)\mbox{Tr}\{ cd\cdots ef\hat{\cal D}\}+\nonumber \\
 & + & \epsilon\sum_{j}\Bigg(\frac{[\eta_{j},c]_\epsilon u_{ij}^{*}}{(1-\epsilon
 e^{-\beta
 E_{j}})}-\frac{[\eta_{\bar{j}}^{\dagger},c]_\epsilon v_{ij}^{*}}{(1-\epsilon e^{\beta
 E_{j}})}\Bigg)\mbox{Tr}\{ bd\cdots ef\hat{\cal D}\}+\cdots\nonumber \\
 & + & \sum_{j}\Bigg(\frac{[\eta_{j},f]_\epsilon u_{ij}^{*}}{(1-\epsilon
 e^{-\beta
 E_{j}})}-\frac{[\eta_{\bar{j}}^{\dagger},f]_\epsilon v_{ij}^{*}}{(1-\epsilon e^{\beta
 E_{j}})}\Bigg)\mbox{Tr}\{ bcd\cdots e\hat{\cal D}\}.
\label{gau11}
\end{eqnarray}

Then, we show the contracted terms is 
\begin{eqnarray}
\widehat{ab}=
\cases{
\sum_{j}\Bigg(\frac{u_{ij}[\eta_{j}^{\dagger},b]_\epsilon}{(1-\epsilon
    e^{\beta{E_{j}}})}-\frac{v_{ij}[\eta_{\bar{j}},b]_\epsilon}{(1-\epsilon
    e^{-\beta{E_{j}}})}\Bigg)\quad\mbox{if}\quad
    a=\alpha_{i}^{\dagger}\cr
    \sum_{j}\Bigg(\frac{[\eta_{j},b]_\epsilon u_{ji}^{*}}{(1-\epsilon
    e^{-\beta{E_{j}}})}-\frac{[\eta_{\bar{j}}^{\dagger},b]_\epsilon v_{ji}^{*}}{(1-\epsilon
    e^{\beta{E_{j}}})}\Bigg)\quad\mbox{if}\quad a=\alpha_{i}}
\label{gau12}
\end{eqnarray}
and, therefore, we write the ensemble average as 
\begin{eqnarray}
\mbox{Tr}\{ abc\cdots ef\hat{\cal D}\} & = & \widehat{ab}\mbox{Tr}\{
cd\cdots ef\hat{\cal D}\}+\epsilon\widehat{ac}\mbox{Tr}\{ bd\cdots
ef\hat{\cal D}\}+\nonumber \\
 & + & \widehat{ad}\mbox{Tr}\{ bc\cdots ef\hat{\cal D}\}+\cdots+\widehat{
 af}\mbox{Tr}\{ bcd\cdots e\hat{\cal D}\}
\label{gau13}
\end{eqnarray}
and, thus, we demonstrated the generalized Wick's theorem at finite temperature using the quadratic Hamiltonian whose
the contracted terms are given in (\ref{gau12}) and depend on the transformation parameters.

\section{An illustrative theorem application}
\label{session3}

To illustrate the utility of the generalized Wick's theorem, we evaluate, in the homogeneous weakly interacting Bose gas at low temperature, the parameter $Q$ define by,
\begin{eqnarray}
Q=\frac{\sqrt{\langle \hat{N}^2\rangle-\langle \hat{N}\rangle^2}}{\langle \hat{N}\rangle}
\label{5}
\end{eqnarray}
which is the measure to the departure of the particle number distribution from a Poissonian distribution \cite{man82}.

For the ideal Bose gas, the fluctuation of the particle number,
at temperature below the the Bose-Einstein condensate transition, have been studied considering the gas inside a
box \cite{zif77} or in harmonic trap \cite{gaj97}. In both, it is shown that microcanonical fluctuations below the critical temperature
tend to zero and scale with the number of particles as $1/\sqrt{\langle\hat{N}\rangle}$. However,
in the case of the weakly interacting gas, it is necessary consider the Hamiltonian with interacting terms.

We evaluate the ensemble average of product of four boson operators,
$a$, $b$, $c$ e $d$. Using (\ref{gau13}) we have 
\begin{eqnarray}
\mbox{Tr}\{ abcd{\cal D}\}=\widehat{ab}\mbox{Tr}\{ cd{\cal D}\}+\widehat{
ac}\mbox{Tr}\{ bd{\cal D}\}+\widehat{ad}\mbox{Tr}\{ bc{\cal D}\},
\label{gau14}
\end{eqnarray}
whereas 
\begin{eqnarray}
\mbox{Tr}\{ abcd{\cal D}\}=\mbox{Tr}\{ ab{\cal D}\}\mbox{Tr}\{
cd{\cal D}\}+\mbox{Tr}\{ ac{\cal D}\}\mbox{Tr}\{ bd{\cal D}\}+\mbox{Tr}\{
ad{\cal D}\}\mbox{Tr}\{ bc{\cal D}\},
\label{gau15}
\end{eqnarray}
we show the expansion of the four operator ensemble average in the sum of the two operator ensemble average 
through the Wick's theorem.

Thus the evaluation of the parameter $Q$ at low temperature regime depends on the fluctuation ensemble average, $\langle \hat{N}^{2}\rangle$,  and the particle number ensemble average, $\langle\hat{N}\rangle$. 
Expanding the creation and annihilation operator in a basis of the plane wave functions with momentum $\vec{k}$, the fluctuation is,
\begin{equation}
\langle \hat{N}^{2}\rangle=\sum_{\vec{k},\vec{k}'}\mbox{Tr}\{
a_{\vec{k}}^{\dagger}a_{\vec{k}}a_{\vec{k}'}^{\dagger}a_{\vec{k}'}{\cal D}\},
\label{num7}
\end{equation}
and the particle number ensemble average is,
\begin{equation}
\langle \hat{N}\rangle=\sum_{\vec{k}}\mbox{Tr}\{
a_{\vec{k}}^{\dagger}a_{\vec{k}}{\cal D}\}.
\label{num7a}
\end{equation}

When the temperature is below to transition,
the Hamiltonian can be approached to a quadratic and the vacuum can
be represented by a coherent state characterized by complex number
$z_{0}$. This vacuum can be described by a shift in particle operator 
\begin{eqnarray}
\cases{
a_{\vec{k}}=c_{\vec{k}}+z_{0}\delta_{\vec{k},0}\cr
a_{\vec{k}}^{\dagger}=c_{\vec{k}}^{\dagger}+z_{0}\delta_{\vec{k},0}}
\label{num8}
\end{eqnarray}
then, using the generalized Wick's theorem, we determine the average values
of shifted operators.

The fluctuation can be written as, 
\begin{eqnarray}
\langle \hat{N}^{2}\rangle & = & \sum_{\vec{k},\vec{k}'}\langle
c_{\vec{k}}^{\dagger}c_{\vec{k}}c_{\vec{k}'}^{\dagger}c_{\vec{k}'}\rangle+2z_{0}^{2}\sum_{\vec{k}}\langle
c_{\vec{k}}^{\dagger}c_{\vec{k}}\rangle+
 \nonumber \\
 & + & z_{0}^{2}(\langle c_{0}^{\dagger}c_{0}^{\dagger}\rangle+\langle
c_{0}c_{0}^{\dagger}\rangle+\langle
c_{0}^{\dagger}c_{0}\rangle+\langle
c_{0}c_{0}\rangle)+z_{0}^{4}.
\label{num9}
\end{eqnarray}
Using the generalized Wick's theorem, 
we obtain
\begin{eqnarray}
\langle \hat{N}^{2}\rangle & = & \sum_{\vec{k},\vec{k}'}(\langle
c_{\vec{k}}^{\dagger}c_{\vec{k}}\rangle\langle
c_{\vec{k}'}^{\dagger}c_{\vec{k}'}\rangle+\langle
c_{\vec{k}}^{\dagger}c_{\vec{k}'}^{\dagger}\rangle\langle
c_{\vec{k}}c_{\vec{k}'}\rangle+\langle c_{\vec{k}}^{\dagger}c_{\vec{k}'}\rangle\langle
c_{\vec{k}}c_{\vec{k}'}^{\dagger}\rangle)+\nonumber \\
 & + & 2z_{0}^{2}\sum_{k}\langle
c_{\vec{k}}^{\dagger}c_{\vec{k}}\rangle+z_{0}^{2}(1+
2\langle c_{0}^{\dagger}c_{0}\rangle+\langle
c_{0}^{\dagger}c_{0}^{\dagger}\rangle+\langle
c_{0}c_{0}\rangle)+z_{0}^{4}.
\label{num11}
\end{eqnarray}

Taking the Bogoliubov transformation,
\begin{equation}
\label{newbog}
c_{\vec{k}}=\cosh(\sigma_{\vec{k}})b_{\vec{k}}-\sinh(\sigma_{\vec{k}})b_{-\vec{k}}^{\dagger}
\end{equation}
where the hyper-geometric angle $\sigma_{\vec{k}}$ is adjusted to diagonalize the quadratic Hamiltonian (\ref{2}).
We have the contracted terms as function of the hyper-geometric angle,
\begin{eqnarray}
\label{gau12a}
\cases{\langle c_{\vec{k}}c_{\vec{k}'}\rangle=\frac{1}{2}\left[\coth\left(\frac{\beta E_k}{2}\right)(1+\sinh(2\sigma_{\vec{k}}))\right]\delta_{\vec{k},-\vec{k}'}\cr
\langle
c^\dagger_{\vec{k}}c_{\vec{k}'}\rangle=\frac{1}{2}\left[\coth\left(\frac{\beta E_k}{2}\right)\sinh(2\sigma_{\vec{k}})-1\right]\delta_{\vec{k},\vec{k}'}},
\end{eqnarray}
where $E_k$ is the low-lying excitation energy. 

Hence, the fluctuation can be written as function of tthe excitation energy, temperature and hyper-geometric angle, 
\begin{eqnarray}
\langle \hat{N}^{2}\rangle & = &
\left(z_{0}^{2}+\frac{1}{2}\sum_{k}[\coth\left(\frac{\beta E_k}{2}\right)\sinh(2\sigma_{\vec{k}})-1]\right)^{2}+
\nonumber
\\
& + & \frac{1}{4}\sum_{k}\left[\coth\left(\frac{\beta E_k}{2}\right)(1+\sinh(2\sigma_{\vec{k}}))\right]^2+
\nonumber\\
&+&\frac{1}{4}\sum_{\vec{k}}\left[\coth\left(\frac{\beta E_k}{2}\right)\sinh(2\sigma_{\vec{k}})-1\right]^2+
\nonumber\\
&+&z_{0}^{2}\lim_{\vec{k}\to 0}\left[\coth\left(\frac{\beta E_k}{2}\right)\sinh(2\sigma_{\vec{k}})\right].
\label{num13}
\end{eqnarray}

In the cases of the fluctuation convergence and at zero temperature the expression (\ref{num13}) resume to,
\begin{eqnarray}
\langle\hat{N}^{2}\rangle=
\frac{1}{2}\sum_{\vec{k}}(\sinh{2\sigma_{\vec{k}}})^{2}+z_{0}^{2}e^{-2\sigma_{0}}.
\label{num15}
\end{eqnarray}

Considering $\sigma_{0}=0$ and knowing that
$(\sinh{\sigma_{\vec{k}}})^{2}\leq\frac{(\sinh{2\sigma_{\vec{k}}})^{2}}{2}$ we can
define a minimum limit to $\langle \hat{N}^2\rangle$ 
and, therefore
\begin{eqnarray}
Q\geq\frac{1}{\sqrt{\langle
    \hat{N}\rangle}}.
\label{num16}
\end{eqnarray}

Although the vacuum is a coherent state, the fundamental state show to be narrower than the Poissonian distribution due to the $Q$ , for high particle number, goes to zero slower than the Poisson distribution.

\section{Conclusion}
\label{session4}

We demonstrated the generalized Wick's theorem at finite temperature considering
a quadratic Hamiltonian and we illustrated this result with a application
to evaluate the fluctuation particle number of a weakly interacting Bose
gas below the condensate transition temperature. 

Different from the reference \cite{gau60}, we prove the generalized Wick's theorem for a quadratic Hamiltonian is similar to the simplified Gaudin's demonstration. However, the contracted terms expressions is a linear combination of to the Bogoliubov parameters weighted by the commutation rule and the statistical distribution which depends on the symmetry of the Fock states.    

An illustrative example was done for determining the parameter $Q$ measure how narrow is the particle distribution. The result
showed the sub-Poissonian character of the distribution due to the Bogoliubov parameters. An appropriate explanation for this behavior will be shown in a future work.

\begin{acknowledgements}
I am grateful to the Prof. Walter Wreszinski for introducing me to the subject.
\end{acknowledgements}

\end{document}